\documentstyle[preprint,aps,graphicx,floats]{revtex}
\def\ba{\begin{eqnarray}}
\def\ea{\end{eqnarray}}
\def\bq{\begin{equation}}
\def\eq{\end{equation}}

\def\sla#1{\ifmmode%
\setbox0=\hbox{$#1$}%
\setbox1=\hbox to\wd0{\hss$/$\hss}\else%
\setbox0=\hbox{#1}%
\setbox1=\hbox to\wd0{\hss/\hss}\fi%
#1\hskip-\wd0\box1 }

\newcommand{\qorqbar}{^{\scriptscriptstyle (}\overline{q}^{\scriptscriptstyle )}}
\begin{document}
\thispagestyle{empty}

\renewcommand{\small}{\normalsize} 

\preprint{
\font\fortssbx=cmssbx10 scaled \magstep2
\hbox to \hsize{
\hskip.5in \raise.1in\hbox{\fortssbx University of Wisconsin - Madison}
\hfill\vtop{\hbox{\bf MADPH-01-1205}
            \hbox{July 2001}} }
}

\title{\vspace{.45in}
Finite-Width Effects in Top Quark Production at Hadron Colliders
}
\author{N.~Kauer and D.~Zeppenfeld\\[3mm]}
\address{
Department of Physics, University of Wisconsin, Madison, WI 53706, USA
}
\maketitle
\begin{abstract}
Production cross sections for $t\bar t$ and $t\bar tj$ events at hadron 
colliders are calculated, including finite width effects and off resonance
contributions for the entire decay chain, $t\to bW\to b\ell\nu$, for both
top quarks. Resulting background rates to Higgs search at the CERN LHC
are updated for inclusive $H\to WW$ studies and for $H\to\tau\tau$ and
$H\to WW$ decays in weak boson fusion events. Finite width effects are large,
increasing $t\bar t(j)$ rates by 20\% or more, after typical cuts 
which are employed for top-background rejection.
\end{abstract}

\vspace{0.2in}


\section{Introduction}
\label{sec:intro}

$t\bar t$ production~\cite{CDFD0} 
is a copious source of $W$-pairs and, hence, of
isolated leptons at the Tevatron and the LHC. Top quark production will
be intensely studied as a signal at these colliders. In addition, it
constitutes an important background for many new particle searches. Examples
include the leptonic signals for cascade decays of supersymmetric 
particles~\cite{SUSY} or searches for 
$H\to W^+W^-$~\cite{DittDrein,CMS,ATLAS_TDR_2,Trefzger_Higgs,KPRZ,RZ_WW} 
and $H\to\tau\tau$~\cite{ATLAS_TDR_2,RZH_tau,PRZ_TauTau} decays. 

\begin{figure}[thb]
\vspace{0.5cm}
\begin{center}
\includegraphics[width=16.5cm]{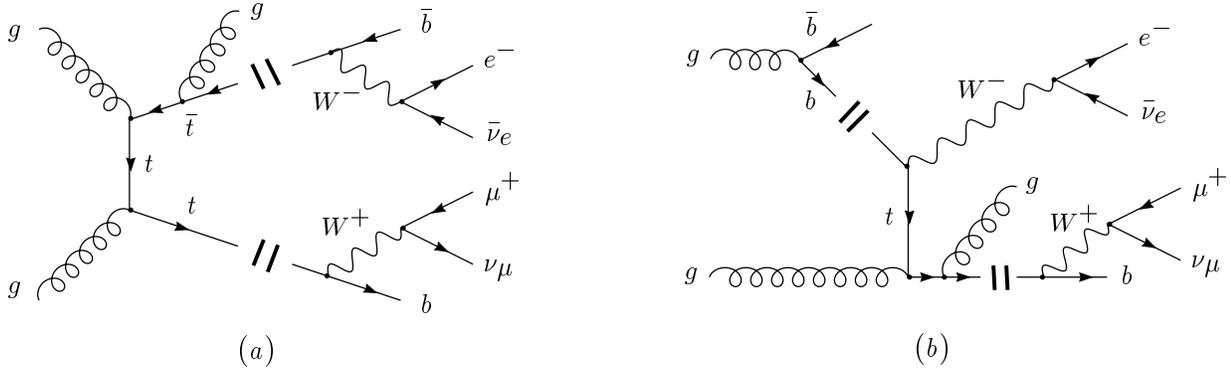} \\
\end{center}
\caption{Feynman diagrams contributing to $gg\to b\bar b
e^-\bar\nu_e\mu^+\nu_\mu g$ in the narrow-width approximation: (a) on-shell 
$t\bar t$ production and (b) $Wt$ single top production. The double bars
indicate heavy quark propagators which may be treated as on-shell particles in
various approximations.}
\vspace{0.5cm}
\label{fig:nwa}
\end{figure}

Usually, $t\bar t$ production is considered in the narrow-width
approximation (NWA), which effectively decouples top production and
decay (see Fig.~\ref{fig:nwa}(a)). Whenever resonant top production dominates,
this approximation is well motivated and it greatly simplifies matrix 
element evaluation and phase space generation.  The NWA is also useful for 
single-resonant top production as shown in 
Fig.~\ref{fig:nwa}(b)~\cite{Boos_SingleTop,Tait_tW,Boos_SingleTop_Followup,OtherSingleTopRefs}.
In some cases calculations have been further simplified by also treating
the decaying $W$ bosons as on-shell particles.  

Naturally, the accuracy of these approximations needs to be tested, which 
requires a full calculation of off-shell effects. Restrictive selection 
cuts, as used for efficient suppression of $t\bar t$ 
backgrounds~\cite{Trefzger_Higgs,RZ_WW}, tend to 
be optimized against on-shell top production
which may substantially enhance the relative importance of off-shell 
contributions. In applying the NWA, another problem inadvertently arises: the
Feynman diagrams in Figs.~\ref{fig:nwa}(a) and \ref{fig:nwa}(b) have identical
initial and final states.  When approximate cross sections, each specialized 
to a particular phase space region, are added up to obtain the total
rate, double-counting can occur, and interference effects in overlap
regions may not be properly accounted for. One thus needs a calculation
which includes both resonant and non-resonant contributions, using finite
width top-quark propagators, which correctly includes interference effects
between the various contributions. The purpose of this paper is to present 
such a calculation for $t\bar t$ and $t\bar tj$ production. 
In addition to merging resonant and non-resonant effects for the top quarks,
we also include finite width effects for the $W$ bosons, i.e. we consider
general $b\bar b e^-\bar\nu_e\mu^+\nu_\mu(j)$ final states. Finite
top width effects, at the level of $t\to bW$ decays, have been considered 
previously for $tWb$ production~\cite{Tait_tW}. The complete set of 
Feynman graphs for $pp\to bW^+\bar bW^-$ processes has been generated as
well~\cite{Boos_SingleTop_Followup}. However, the full treatment of 
lepton final states with spin correlations and off-shell contributions is 
new even for the $t\bar t$ case. To our knowledge, 
no previous calculation of finite width effects in $t\bar tj$ production 
exists.

The paper is organized as follows. In Section~\ref{sec:method} we discuss
various methods of including finite width effects and discuss their 
advantages and their practicality. We adopt the ``overall
factor scheme'' and apply it in Section~\ref{sec:general} to 
$b\bar b e^-\bar\nu_e\mu^+\nu_\mu$ production and the analogous process with 
one additional colored parton in the final state. While the matrix elements 
can, in principle, be generated with automated programs like
MADGRAPH \cite{MADGRAPH}, 
the proper inclusion of finite widths, preservation of electroweak and
strong gauge invariance, avoidance of double counting and of divergences
in extraneous phase space regions requires manual intervention.
In Section~\ref{sec:general} we describe the content of our calculation,
the various consistency tests, and other important features of the program
which we have developed. 

Our program has already been used to study backgrounds to $H\to WW$ decays
at the LHC~\cite{KPRZ}. We expand on this analysis in 
Section~\ref{sec:application} and use $H\to WW$ and $H\to\tau\tau$ decays, 
more precisely
the backgrounds produced by $t\bar t$ and $t\bar tj$ production,
to exemplify the size of off-shell and on-shell contributions and compare 
our full simulations with previous background estimates. A summary and 
conclusions are given in Section~\ref{sec:summary}.


\section{Finite-width effects and gauge invariance}
\label{sec:method}

The inclusion of finite width effects is needed in order to avoid the singular
behavior of the tree level propagators on mass-shell, $p^2-m^2=0$.
An approach that is straightforward to implement and, hence, well suited for
automatic code generators like MADGRAPH/HELAS \cite{MADGRAPH,HELAS} is to
use a Breit-Wigner-type propagator with fixed width for all top and $W$
propagators, making no distinction between time-like and space-like 
momenta. For massive fermions like the top quark one simply
substitutes
\bq 
\label{eq:tprop_naive}
\sla{S}(p) = \frac{i}{\sla{p} - m} = \frac{i(\sla{p} + m)}{p^2 - m^2} \ 
    \longrightarrow \ \frac{i(\sla{p} + m)}{p^2 - m^2 + i m \Gamma}\;.
\eq

However, this technique, also referred to as {\em fixed-width scheme}, will
lead to gauge-dependent matrix elements. Gauge invariance requires that the 
Ward identity 
\bq 
k_\mu V^\mu = -i g \; [i\sla{S}^{-1}(p_1) - i\sla{S}^{-1}(p_2)]
\label{eq:ward-id}
\eq
(with $p_1=k+p_2$) be satisfied, where $V^\mu = -i g \gamma^\mu$ is the 
gauge boson -- fermion vertex. For the top-propagator of 
Eq.~(\ref{eq:tprop_naive}), the inverse, $\sla{S}^{-1}$, even diverges at 
$p^2=m^2$ and the Ward identity is violated. The naive use of Breit-Wigner
propagators does not produce consistent matrix elements.

Calculating vertex functions and inverse propagators perturbatively, the 
Ward identity of Eq.~(\ref{eq:ward-id}) will be satisfied order by order.
This is the basis of the so called {\em fermion loop scheme} for the 
$W$-propagator where, for a LO calculation, the imaginary part of the 
fermionic 1-loop corrections is included in both the vertex function and in 
the inverse propagator~\cite{Baur:1995aa,Argyres:1995ym,Beenakker:1997kn}. 
For the propagator this corresponds to the Dyson resummation of the 
imaginary part of the $W$ vacuum polarization.

\begin{figure}[tb]
\vspace{0.5cm}
\begin{center}
\includegraphics[width=16.5cm]{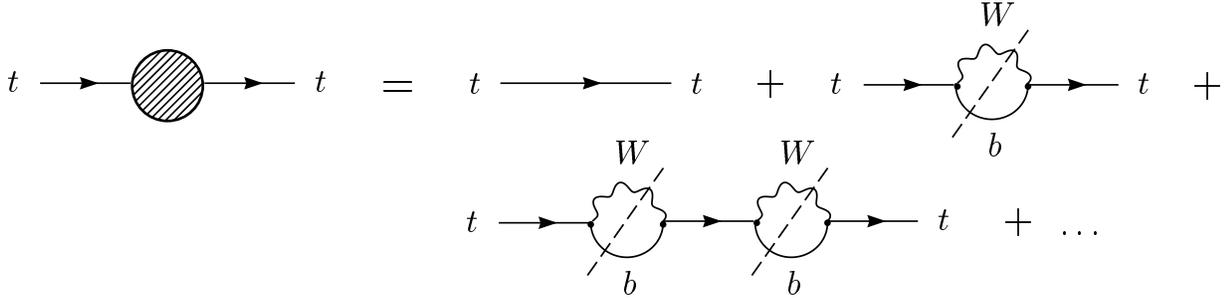} \\
\end{center}
\caption{Dyson resummation of the imaginary part of the $Wb$
contribution to the top self-energy.  The resummed top propagator is
represented by the cross-hatched blob.  
}
\vspace{0.5cm}
\label{fig:topprop}
\end{figure}

\begin{figure}[thb]
\vspace{0.5cm}
\begin{center}
\includegraphics[width=16.5cm]{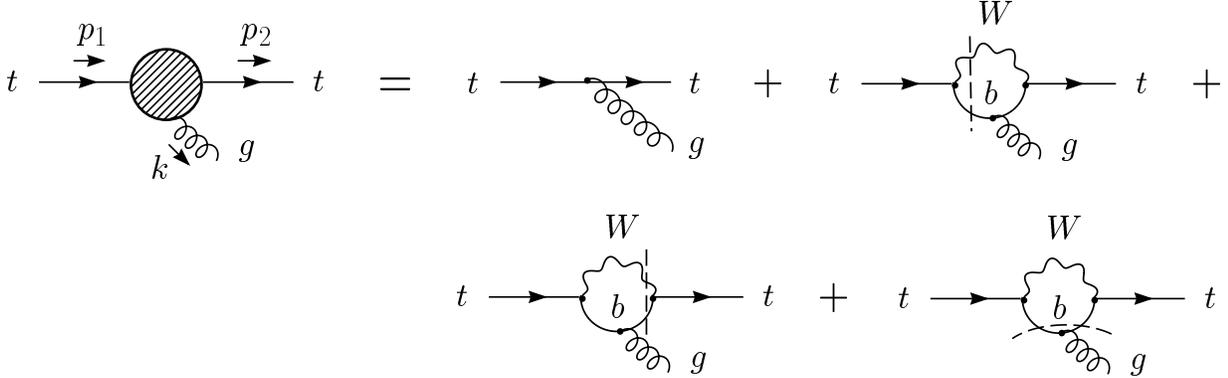} \\
\end{center}
\caption{Effective $ttg$ vertex including the imaginary part of the 
$Wb$ loop-correction to the
tree-level vertex.  The imaginary part is obtained by cutting the triangle
graph in all possible ways corresponding to on-shell intermediate states.}
\vspace{0.5cm}
\label{fig:ttgvert}
\end{figure}

A theory driven solution of the finite width problem for the top quark
propagator would generalize this scheme. More specifically, a
Dyson resummation of the imaginary parts of the 1-loop self-energy of the
top quark, due to $bW$ intermediate states (see Fig.~\ref{fig:topprop}),
results in the effective propagator
\bq
\sla S_t(p) = \frac{i}{\sla{p} - m_t + i \gamma(p^2) \sla{p} P_L \;
\theta(p^2-(m_W+m_b)^2)} 
\label{eq:topprop}
\eq
Here $P_L$ is the left-chiral projector and 
\begin{displaymath}
\textstyle
\gamma(p^2) =
\frac{1}{64\pi}\;\frac{g^2}{m_W^2}\;\sqrt{\left[1 - 
\frac{(m_W + m_b)^2}{p^2}\right]\left[1 - \frac{(m_W - m_b)^2}{p^2}\right]}\
\left[\left(1-\frac{m_W^2}{p^2}\right)(2m_W^2+p^2)+\frac{m_b^2}
{p^2}(m_W^2+m_b^2-2p^2)\right]
\end{displaymath}
In order to satisfy the $SU(3)$ Ward identity of Eq.~(\ref{eq:ward-id})
one also needs to calculate the imaginary part of the $ttg$ vertex 
(see Fig.~\ref{fig:ttgvert}). We have checked by explicit calculation 
that this $SU(3)$ Ward identity is indeed satisfied. However, the effective 
$ttg$ vertex already is too complex to be displayed 
here.\footnote{It should be noted that the simplifications that
  led to concise results for the effective $WW\gamma$ and $WW\gamma\gamma$
  vertices (see Refs.~\protect\cite{Baur:1995aa,Kauer_Master,Baur:1997bn})
  can not be applied when evaluating the effective $ttg$ vertex.}  
For our applications we would need to know the imaginary contributions 
to $ttgg$ and $ttggg$ vertices as well.

\begin{figure}[bht]
\vspace{0.5cm}
\begin{center}
\includegraphics[width=12.0cm]{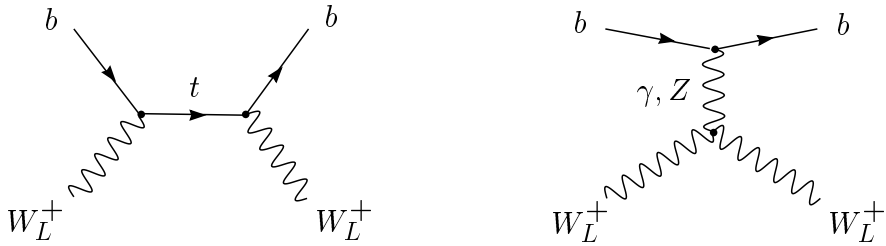} \\
\end{center}
\caption{The electroweak gauge cancellation in $b W^+_L\to b W^+_L$ is 
perturbed by the resummed one-loop approach.}
\vspace{0.5cm}
\label{fig:bbWW}
\end{figure}

A further complication arises from the need of electroweak gauge 
invariance in our calculations. Consider the simple process (or subgraph)
depicted in Fig.~\ref{fig:bbWW}: elastic scattering of a $b$-quark and
a longitudinal $W$ boson. The longitudinal polarization vectors of the two
$W$'s scale like $\sqrt{\hat s}/m_W$, which leads to a rise of the two 
subamplitudes with the center of mass energy:
\ba
{\cal M}_t &\sim & {\hat s^2\left[ 1+ i \gamma(\hat s)\right]
\over \hat s \left[1+ i \gamma(\hat s)\right]^2 - m_t^2} 
\nonumber \\ 
{\cal M}_{\gamma,Z} &\sim & {\hat s^2\over \hat t - m_V^2 }\; .
\ea
As is well known for the crossed process $e^+e^-\to W^+W^-$, gauge invariance 
of the electroweak couplings leads to a cancellation of these leading terms
and results in partial wave amplitudes which do not grow with energy and,
thus, respect partial wave unitarity~\cite{HPZH}. When using the finite
width top-quark propagator of Eq.~(\ref{eq:topprop}), with 
$\gamma(\hat s)\sim \hat s$ at high energy, the width correction dominates
the propagator at very high $\hat s$ and leads to 
${\cal M}_t\sim {\rm const}$ at high energy, thus spoiling the gauge theory 
cancellations between ${\cal M}_t$ and ${\cal M}_{\gamma,Z}$: the $bW_L\to
bW_L$ scattering amplitude violates unitarity in the $J={1\over 2}$ partial 
wave at sufficiently high energy. The likely solution to this problem lies
in adding the imaginary parts of $btW$, $btWg$ vertices etc. to the loop 
scheme prescription, a solution which clearly becomes too involved for 
practical applications.

When considering cross sections which contain contributions from both
resonant and non-resonant amplitudes,
\bq 
\sigma \sim \int d\,{\rm PS}_{\scriptstyle i\to f} \ 
    |{\cal M}_{\scriptstyle if, res. }
     + {\cal M}_{\scriptstyle if, nonres. } |^2 \;,
\eq
we need an alternative which guarantees gauge invariance and unitarity while 
allowing the effective substitution of propagators by a Breit-Wigner 
form in the resonant contributions, ${\cal M}_{\scriptstyle if, res}$. 
In our work, we adopt the overall factor 
scheme~\cite{bvz} which starts from the observation that the zero width
amplitudes, as derived from unresummed Feynman rules, provide a gauge 
invariant expression with proper high energy behavior. 
Multiplying the full lowest order amplitude by an overall factor 
$(p^2-m^2)/(p^2-m^2+im\Gamma)$, for each resonant propagator, preserves gauge
invariance but effectively replaces the tree level propagators, which are 
divergent on mass-shell, by finite-width Breit-Wigner propagators,
\bq 
{\cal M}_{\scriptstyle if \; (\Gamma = 0)} \ 
    \frac{p^2 - m^2}{p^2 - m^2 + i m \Gamma}
    = {\cal M}_{\scriptstyle if, res. \; (\Gamma \not= 0)}
    + {\cal M}_{\scriptstyle if, nonres. \; (\Gamma = 0)} \ 
    \frac{p^2 - m^2}{p^2 - m^2 + i m \Gamma}
\eq

In the overall factor scheme, the cross section is thus calculated in terms 
of this modified amplitude. 
Note that the additional overall factor is close to $1$ in
phase space regions where the non-resonant amplitudes yield significant
contributions. Close to resonance, on the other hand, the amplitude 
effectively reduces to the dominant resonant terms. 
As we will show in Section~III, this approach provides an excellent
interpolation between on and off resonance regions, with ambiguities never
rising beyond the 1--2\% level.

A second practical solution starts from the observation that
the Ward identity in Eq.~(\ref{eq:ward-id}) remains fulfilled if 
one changes constant terms in the inverse propagator. 
This suggests another method to restore gauge invariance, which
has been dubbed the {\em complex-mass scheme\/} \cite{Denner:1999gp,compMother}.
Its finite-width amplitude is derived from the
full lowest-order amplitude with zero-width propagators by substituting all
$W,$ $Z$ and top quark masses according to
\bq 
m \longrightarrow \sqrt{m^2 - i m \Gamma} 
\label{eq:complexmass}
\eq
This scheme has recently been used in a single top quark study for the
Tevatron \cite{vanderHeide:2000fx}. An unphysical consequence of the
universal substitution (\ref{eq:complexmass}) 
is that space-like propagators receive imaginary contributions, or that 
the top-Higgs Yukawa coupling, $h_t=m_t/v$, receives 
an imaginary part. However, such effects are suppressed by factors 
$\Gamma/m \ll 1$ and would presumably not be noticeable in a LO Monte Carlo
program. 

In our finite width $t\bar t$ and $t\bar tj$ Monte Carlo programs,
we have opted for the overall factor scheme.  We expect very similar 
numerical results in the complex-mass scheme, which we have not implemented, 
however.


\section{Cross Sections for \lowercase{$bb$}$WW$\lowercase{$(j)$} production}
\label{sec:general}

Unstable particles occur in several places in top quark pair production 
processes: in the form of decaying top-quarks and also as the $W$ bosons 
which arise in their decays. In order to include off-resonance contributions 
for both, we are led to consider the full tree-level matrix elements for 
$b\bar b e^-\bar\nu_e\mu^+\nu_\mu$ final states (``$t\bar t$ production'') at 
${\cal O}(\alpha_s^2\alpha^4)$, and final states with one additional
parton (``$t\bar tj$ production'') at ${\cal O}(\alpha_s^3\alpha^4)$.

\subsection{Matrix Elements}
\label{subsec:matrixel}

For $pp$ or $p\bar p$ collisions, matrix elements for the
following subprocesses need to be evaluated (we neglect CKM mixing):

\ba
& & gg \to b\bar b e^-\bar\nu_e\mu^+\nu_\mu \;, \qquad 
q\bar q \to b\bar b e^-\bar\nu_e\mu^+\nu_\mu \;\qquad \mbox{for}\ 
t\bar t  \label{eq:tt-mat} \\
& & gg \to b\bar b e^-\bar\nu_e\mu^+\nu_\mu g\;, 
\qquad g\qorqbar \to b\bar b e^-\bar\nu_e\mu^+\nu_\mu \qorqbar\;, \qquad  
q\bar q \to b\bar b e^-\bar\nu_e\mu^+\nu_\mu g \qquad \mbox{for}\ t\bar tj \;.
\label{eq:ttj-mat} 
\ea

\begin{figure}[thb]
\vspace{0.5cm}
\begin{center}
\includegraphics[width=16.5cm]{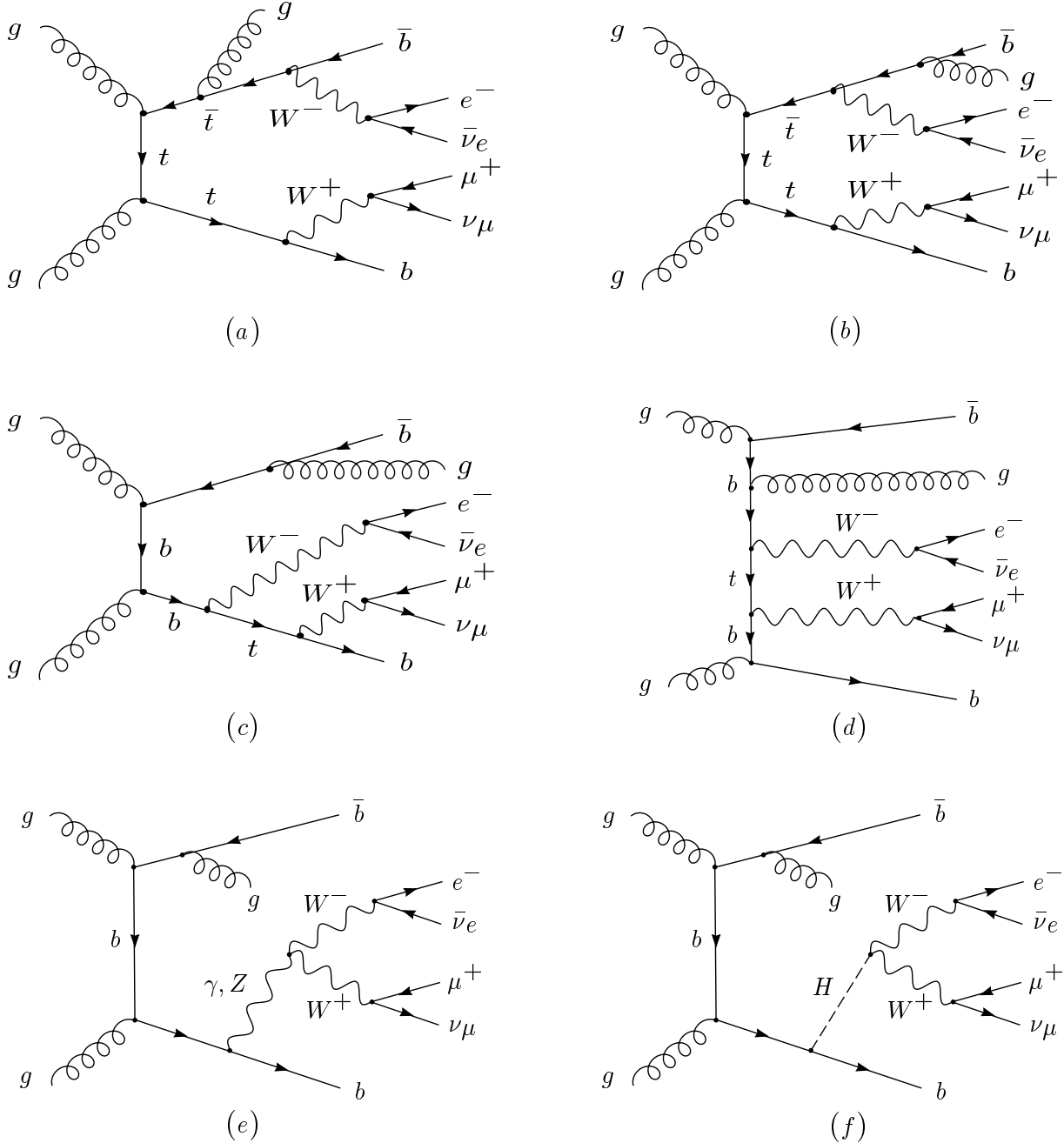} \\
\end{center}
\caption{Feynman diagrams contributing to $gg\to b\bar b
e^-\bar\nu_e\mu^+\nu_\mu g$ with off-shell intermediate states:  
double-resonant (a,b), single-resonant (c) and non-resonant (d,e,f)
contributions.}
\vspace{0.5cm}
\label{fig:offshell}
\end{figure}

\begin{figure}[thb]
\vspace{0.5cm}
\begin{center}
\includegraphics[width=15.0cm]{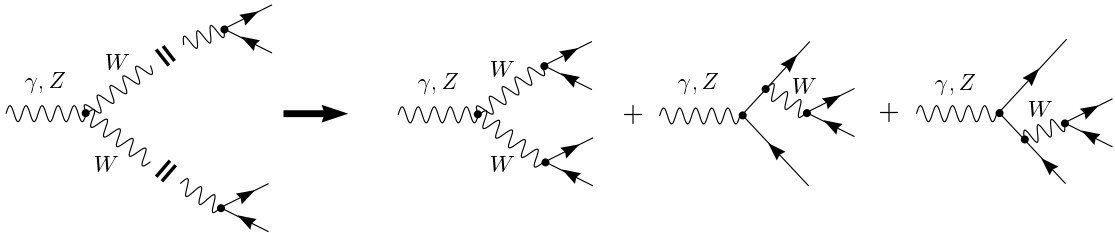} \\
\end{center}
\caption{Amplitude factorization for $\gamma$ and $Z$ decays.
Explicit summation of the sketched helicity amplitude fragments 
leads to a significant computational reduction,
 as shown in
Table~\protect\ref{tab:amps}.}
\vspace{0.5cm}
\label{fig:ampsum}
\end{figure}

Representative Feynman graphs
are shown in Fig.~\ref{fig:offshell}. Double resonant contributions 
include gluon radiation off initial state partons, the top quarks,
and final state $b$-quarks (Figs.~\ref{fig:offshell}(a) and (b)).
An example for a single resonant graph is shown in Fig.~\ref{fig:offshell}(c),
while (d) depicts one of the non-resonant graphs. Electroweak gauge invariance
of the $b\bar b\to W^+W^-$ subgraphs in (c) and (d) requires inclusion of 
$\gamma$ and $Z$ exchange contributions. One such contribution is shown in 
Fig.~\ref{fig:offshell}(e). Others include $W$-emission off the final state
lepton lines (see Fig.~\ref{fig:ampsum} and discussion below). Our code 
includes finite $b$-quark masses (set to a default value of $m_b=5$~GeV).
This allows to integrate over the entire $b$-quark phase space, including
the $g\to b\bar b$ splitting region. A finite $b$-quark mass necessitates
new contributions, however, namely Higgs exchange diagrams like the one
depicted in Fig.~\ref{fig:offshell}(f). Our code includes all these 
contributions. We avoid goldstone boson exchange graphs 
by working in the unitary gauge for the electroweak sector.

Our calculation assumes different lepton flavors in the decay
of the two $W$-bosons. However, the amplitudes for this mixed lepton 
flavor case can also be used to 
obtain approximate results for same flavor processes, specifically
$e^-\bar\nu_e e^+\nu_e$ and $\mu^-\bar\nu_\mu\mu^+\nu_\mu$ final states.
The double- and single-resonant contributions (with respect to top) are 
identical for the mixed and same flavor sample. However, the latter also 
features $(\gamma,Z \to \ell^+\ell^-) \circ (Z \to \nu_\ell\bar\nu_\ell)$ 
graphs, non-resonant from the view-point of top-decay, which 
do not occur in the mixed flavor case.  Away from the $Z$-boson 
mass-shell and small $\ell^+\ell^-$ invariant mass, these contributions are
small, and the complete $\ell^\pm_1\ell^\mp_2$ cross section 
(with $\ell_{1,2} =e,\mu$) is obtained by multiplying the result presented
below with a lepton-flavor factor of 4.

\begin{table}
\caption{Number of Feynman graphs contributing to different subprocesses.  
The numbers in brackets reflect the reduction of subamplitudes 
illustrated in Fig.~\protect\ref{fig:ampsum} (see text).  The columns 
correspond to different initial states.}
\vspace{0.15in}
\label{tab:amps}
\begin{tabular}{c|cc|cc|cc}
& \multicolumn{2}{c|}{$gg$} &
\multicolumn{2}{c|}{$gq$} & \multicolumn{2}{c}{$qq$}  \\
\hline
$t\bar t$ & 87 & (39) & \multicolumn{2}{c|}{-} & 40 & (16) \\ 
\hline
$t\bar tj$ & 558 & (258) & 246 & (102) & 246 & (102) 
\end{tabular}
\end{table}

The number of subamplitudes, corresponding to individual Feynman graphs,
is sizable for the processes of Eqs.~(\ref{eq:tt-mat},\ref{eq:ttj-mat}) 
and is listed in Table~\ref{tab:amps}. 
Constructing the code that evaluates the matrix elements was simplified by
using automatically generated output of MADGRAPH \cite{MADGRAPH} as a
starting point.  However, the code had to be modified by hand in several
ways. First, every Feynman graph amplitude needs to be multiplied with the 
proper overall factors 
\bq
f_a(p) = {p^2-m^2_a\over p^2-m^2_a+im_a\Gamma_a}\; ,
\eq 
depending on its resonance structure.  For $t\bar t$
production the factors correspond to a resonant top quark with momentum
$p_b+p_{\mu^+}+p_{\nu_\mu}$ (for which we use the shorthand notation
$p_{b\mu^+\nu_\mu}$) and/or a resonant anti-top quark with momentum 
$p_{\bar b e^-\bar\nu_e}$.  For $t\bar tj$ production two additional factors
$f_t(p_{b\mu^+\nu_\mu j})$ and $f_t(p_{\bar b e^-\bar\nu_e j})$ appear, 
corresponding to gluon emission off the final state $b$ or $\bar b$.
In both cases two
$W$ factors ($f_W(p_{e^-\bar\nu_e})$ and $f_W(p_{\mu^+\nu_\mu})$) and one $Z$
factor $f_Z(p_{e^-\bar\nu_e\mu^+\nu_\mu})$ need to be explicitly
multiplied into amplitudes that are non-resonant with respect to these
propagators.  MADGRAPH generates all top, $W$ and $Z$ propagators
automatically as resonant propagators, i.e.~with a finite fixed width.
These propagators can be viewed as the result of multiplying the zero-width
tree-level propagators with the overall factor, or, alternatively, as obtained 
by the substitution 
\bq
[p]_a: \quad (p^2-m_a^2)^{-1} \to (p^2-m_a^2+im_a\Gamma_a)^{-1}\;, 
\eq
which we denote by the symbol $[p]_a$ in the amplitudes given below.
For the first few Feynman graphs of Fig.~\ref{fig:offshell} these changes 
can be summarized as
\ba
\mbox{a)}& &  \quad {\cal M}([p_{b\mu^+\nu_\mu}]_t, 
[p_{\bar b e^-\bar\nu_e}]_t,
[p_{\bar b e^-\bar\nu_e j}]_t, [p_{e^-\bar\nu_e}]_W, 
[p_{\mu^+\nu_\mu}]_W) * f_t(p_{b\mu^+\nu_\mu j}) 
 f_Z(p_{e^-\bar\nu_e\mu^+\nu_\mu}) \;, \\
\mbox{b)}& &  \quad {\cal M}([p_{b\mu^+\nu_\mu}]_t, 
[p_{\bar b e^-\bar\nu_e j}]_t, [p_{e^-\bar\nu_e}]_W, 
[p_{\mu^+\nu_\mu}]_W) * f_t(p_{\bar b e^-\bar\nu_e}) f_t(p_{b\mu^+\nu_\mu j}) 
 f_Z(p_{e^-\bar\nu_e\mu^+\nu_\mu})  \;, \\
\mbox{c)}& &  \quad {\cal M}([p_{b\mu^+\nu_\mu}]_t, 
[p_{e^-\bar\nu_e}]_W, [p_{\mu^+\nu_\mu}]_W) * 
f_t(p_{\bar b e^-\bar\nu_e})  f_t(p_{b\mu^+\nu_\mu j}) 
f_t(p_{\bar b e^-\bar\nu_e j})  f_Z(p_{e^-\bar\nu_e\mu^+\nu_\mu}) \;,  \\
\mbox{d), f)}& &  \quad {\cal M}([p_{e^-\bar\nu_e}]_W, 
[p_{\mu^+\nu_\mu}]_W) * f_t(p_{b\mu^+\nu_\mu}) 
 f_t(p_{\bar b e^-\bar\nu_e})  f_t(p_{b\mu^+\nu_\mu j})  
f_t(p_{\bar b e^-\bar\nu_e j})  f_Z(p_{e^-\bar\nu_e\mu^+\nu_\mu}) \;,  \\
\mbox{e)} \;(Z)& & \quad {\cal M}([p_{e^-\bar\nu_e}]_W,
[p_{\mu^+\nu_\mu}]_W, [p_{e^-\bar\nu_e\mu^+\nu_\mu}]_Z) * 
f_t(p_{b\mu^+\nu_\mu}) 
f_t(p_{\bar b e^-\bar\nu_e})  f_t(p_{b\mu^+\nu_\mu j})  
f_t(p_{\bar b e^-\bar\nu_e j})\;.
\ea

In the overall factor scheme, imaginary parts for $t$-channel top-quark 
propagators are absent. The top-width was eliminated by hand from the 
MADGRAPH output for these space-like propagators. The effect of this 
modification should be small, however,  since $|p^2-m^2|\gg m\Gamma$ if 
$p^2 < 0,$ due to $\Gamma/m \ll 1$.  
The number of Feynman diagrams for $t\bar tj$ production is formidable, 
partially due to repeating sequences of subgraphs, like the ones depicted 
in Fig.~\ref{fig:ampsum}. These subgraphs were combined to effective 
$\gamma/Z$-currents. As shown in Table~\ref{tab:amps}, this procedure
reduces the number of sub-amplitudes, which need to be calculated, by a 
factor of two or more.\footnote{Code generators have some freedom in the
  composition of basic elements when constructing helicity amplitudes, as
  explained in Section~2.7 of Ref.~\protect\cite{HELAS}.  In rare cases, the
  algorithm employed by MADGRAPH is incompatible with the
  factorization outlined in Fig.~\protect\ref{fig:ampsum}.  In these cases,
  suitable amplitudes were composed by hand.}

\begin{figure}[t]
\vspace{0.5cm}
\begin{center}
\includegraphics[width=12.5cm]{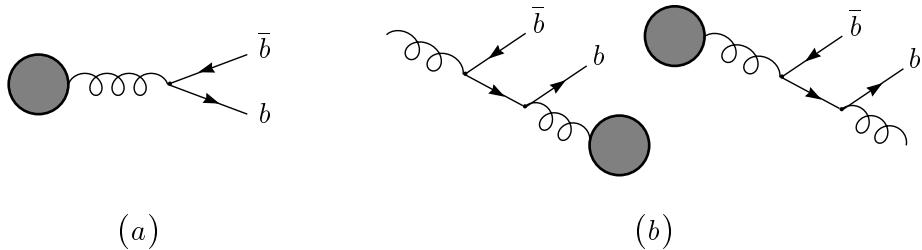} \\
\end{center}
\caption{QCD radiative correction fragments, corresponding to 
(a) $g\to b\bar b$ and (b) $g\to b\bar bg$ splitting. The shaded blob 
represents the rest of the Feynman graph. Care has to be taken to count 
these contributions only once when combining relevant backgrounds.}
\vspace{0.5cm}
\label{fig:hiqcd}
\end{figure}

Since our calculation includes full matrix elements to a high order in 
perturbation theory, care has to be taken to avoid overlap 
with other backgrounds and double-counting. Closer
inspection of the matrix elements for the $t\bar t$ and $t\bar tj$ modes
yields two groups of Feynman diagrams that are candidates for
double-counting.  They are schematically depicted in Fig.~\ref{fig:hiqcd}.
Consider $gu\to b\bar bW^+W^-u$, a subprocess of $t\bar tj$ production,
as an example.  Graph (a) represents $g\to b\bar b$ splitting for a final
state gluon. It constitutes a QCD radiative correction to $gu\to W^+W^-ug$
and should be counted in the QCD $WWjj$ background.  For small $b\bar b$
invariant masses, graph (a) is a contribution to the gluon fragmentation,
which must be counted only once. Group (b) features $g\to b\bar bg$ splitting,
where the $b$-quarks and one gluon are external particles.
When the on-shell gluon is in the initial state and the $\bar b$
is collinear to it, this group represents an $\alpha_s$ correction to 
$bu\to uW^+W^-b$ production. When both the $b$ and the $\bar b$ are collinear
to the initial gluon, we are considering an 
$\alpha_s^2$ QCD correction to $gu\to uW^+W^-$. It is inappropriate to 
include either group as non-resonant contributions to $t\bar tj$ production.
These splitting contributions can be eliminated either through suitable 
cuts or by explicitly excluding the relevant Feynman graphs.  A numerical 
comparison shows that typical selection cuts, like for the Higgs search to be 
considered in Section~\ref{sec:application}, are sufficient to render
the splitting contributions negligible. Since the analyses in
Refs.~\cite{RZ_WW} and \cite{PRZ_TauTau} do not include QCD corrections to
general $WWjj$ production, we chose to omit these graphs in our final
calculations.  This is permissible since the graphs of group (a) and (b)
necessarily contain electroweak interactions of $u$ and $d$ quarks.  
In the context of top backgrounds, the primary focus is on 
electroweak interactions of bottom and top quarks.  
When CKM mixing is neglected, the gauge invariance of our
amplitudes is preserved when setting all electroweak interactions of the
first two quark generations to zero.  This procedure eliminates the graphs 
of groups (a) and (b) and avoids double-counting.

When integrating over collinear $g\to b\bar b$ configurations for initial
state gluons, the
differential cross section receives enhancement factors of order
$\log(\mu_f^2/m_b^2)$.  Since $\alpha_s(m_t)\log(m_t^2/m_b^2)\approx 0.85$, 
one may wonder whether our perturbative leading-order calculations are still
reliable, or whether a resummation of these collinear logarithms is
required.  The correct treatment of the collinear region would include
$gb$ scattering, convoluted with the $b$-quark parton density.  Then, a
subtraction of the gluon splitting term would also be required to avoid
double-counting \cite{Boos_SingleTop,Tait_tW}.  

In the cases at hand, the net effect of the $b$-quark PDF contribution is 
small, however. For inclusive
$pp\to b\bar bWW,$ only 10.4 pb are added to 622 pb (see Table~1 in
Ref.~\cite{Boos_SingleTop_Followup}) .  For the SM Higgs searches via 
weak boson fusion, discussed
in Section~\ref{sec:application}, the tagging criteria and selection cuts are
chosen such that for $t\bar t$ production both $b$-quarks are resolved, 
with $p_T > 20$~GeV. This avoids the collinear region.
For $t\bar t$+1 jet production, the $b$-quark observed as a forward tagging 
jet is required to have $p_T > 20$ GeV, while
the other $b$-quark has no lower transverse momentum threshold.  However,
these collinear regions contribute little to the cross section within
typical cuts. For example, in the $H\to\tau\tau\to e^\pm\mu^\mp\sla{p}_T$ 
search with forward jet tagging cuts~\cite{PRZ_TauTau}, the phase space 
region with $p_T < 20$~GeV for the untagged $b$- (or $\bar b$-) 
quark contributes only 8\% to the total cross section.  

The smallness of these collinear effects is related to the fact that we
are generating jets and $b$-quarks of $p_T>20$~GeV as explicit partons in our
calculations. In order to avoid double counting, the factorization 
scale in the $b$-quark PDF should then be chosen as 
$\mu_f=20$~GeV, which mitigates the role of the
collinear logarithms, $\alpha_s \log(\mu_f^2/m_b^2)$.\footnote{As in any LO
   multi-parton calculation, the dangerous large logarithms are of the form
   $\alpha_s \log (m_t^2/p_{Tj}^2)$. 
   An improvement of our calculation would
   first of all require the resummation of these contributions. Initial state
   collinear logs from $g\to b\bar b$ splitting are unimportant by comparison.}
We conclude that a special treatment of collinear effects is not required in
our studies. We regularize the $b$-quark collinear region by the finite
$b$-quark mass, which provides a simplified but adequate model for the 
$b$-quark PDF. 

Differential cross sections for top production in the narrow-width
approximation are independent of $m_H$.  However, once all off-shell
effects are included, a dependence on $m_H$ is caused by Higgs propagators
that appear in the non-resonant contributions of Fig.~\ref{fig:offshell}(f).
This dependence has a negligible effect on the rates considered in this paper.

\subsection{Phase Space Generator for \lowercase{$t\bar tj$} Production and
Numerical Tests}
\label{subsec:phasespace}

\begin{figure}[thb]
\vspace{0.5cm}
\begin{center}
\includegraphics[width=9.0cm]{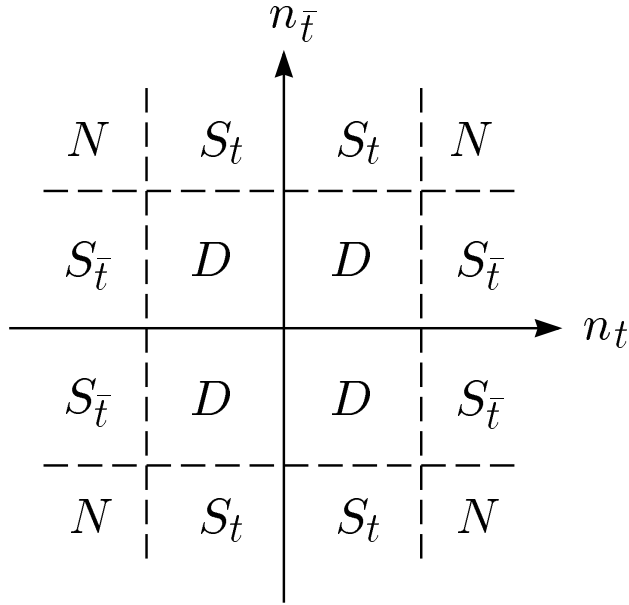} \\
\end{center}

\caption{Phase space decomposition for $bbWW$ final state based on the
variables $n_t = (m_{bW^+}-m_t)/\Gamma_t$ and $n_{\bar t} = (m_{\bar
bW^-}-m_t)/\Gamma_t$.  The calculation presented here employs separate,
hand-optimized phase space generators for double- ($D$), single-
($S_t$ and $S_{\bar t}$) and non-resonant ($N$) phase space regions.  
By default the boundaries are chosen at $|n_{t,\bar t}| = 8$.}

\vspace{0.5cm}
\label{fig:phasespace}
\end{figure}

While a simple phase space generator proved sufficient for $t\bar t$
production, for $t\bar tj$ production with significant selection cuts 
a composite phase space generator that interfaces optimized mappings
for double-, single- and non-resonant phase space regions becomes a
necessity.  These three regions are defined with the help of the two
variables

\bq n_t = \frac{m_{bW^+}-m_t}{\Gamma_t} \qquad \mbox{and} \qquad 
n_{\bar t} = \frac{m_{\bar bW^-}-m_t}{\Gamma_t}
\label{ps-vars}
\eq

The double-resonant region is then defined by
$|n_t|, |n_{\bar t}| < n_c$, 
the single-resonant region by 
$(|n_t| < n_c \wedge |n_{\bar t}| > n_c) \vee (|n_t| > n_c 
\wedge |n_{\bar t}| < n_c)$
and the non-resonant region by 
$|n_t|, |n_{\bar t}| > n_c$ 
(see Fig.~\ref{fig:phasespace}).
In our calculations the boundary parameter $n_c$ is chosen to be 8 by
default (but see tests below).  Since the runtime of the $t\bar tj$ program
is fairly long, the code currently evaluates only the $S_t$ region and
multiplies the result by two to account for $S_{\bar t}$.  This symmetry
assumption can easily be removed in the source code should the need arise.

To assure the correctness of the programs three tests were performed. First,
the Lorentz-invariance of the modified matrix elements is highly sensitive 
to errors.  A suitable test variable can be defined in
the following way: One generates a phase space configuration and evaluates
the matrix element squared, summed over all helicity
combinations.\footnote{ Because of the massive external $b$-quarks, the matrix
element for any particular helicity combination is not boost-invariant.}
Applying an arbitrary boost (with $\gamma < 5$) to all external
momenta and re-evaluating the matrix element, one finally computes

\bq \left|1 - \frac{\sum|{\cal M}_{\scriptstyle
      boosted}|^2}{\sum|{\cal M}|^2}\right|\;.
\label{eq:test-var}
\eq

Evaluating $50\!\cdot\! 10^6$ phase space configurations (within the 
forward jet tagging cuts of Ref.~\protect\cite{PRZ_TauTau}),
we found an average (maximal) test variable of $2\!\cdot\! 10^{-10}$
($7\!\cdot\! 10^{-8}$) for $t\bar t$ and $1\!\cdot\! 10^{-10}$ 
($2\!\cdot\! 10^{-5}$) for $t\bar tj$. \\
The test variable is highly sensitive to errors in the matrix element:  
Omitting the overall factors for a
single amplitude that corresponds to a single-resonant Feynman diagram,
increases the average and maximal test variable by 5-6 orders of
magnitude.
Second, the same test variable was used to check the modifications that 
are necessary to implement the reduction in Fig.~\ref{fig:ampsum}.  

A more recent MADGRAPH version
can be used (with minor modifications) to automatically generate the full
matrix elements of Eqs.~(\ref{eq:tt-mat}) and (\ref{eq:ttj-mat}) with 8 and 9
external particles, respectively.  These matrix elements can then be
compared numerically to the factorized matrix elements in our
programs.  All overall factors need to be set to 1 for this test and
finite widths for space-like top propagators need to be restored.
A single, wrong coupling constant, e.g.~$g_{Zu}$ rather than $g_{Zd}$,
increases the average and maximal test values by 11-13 orders of
magnitude.

Finally, the phase space generation for $t\bar t$ has been tested by
comparing with known cross sections for the Tevatron and the LHC.  The
composite phase space generation for $t\bar tj$ has been checked by moving
the boundary between phase space regions, i.e.~changing the value of $n_c$.
For different cut sets and $n_c =$ 4, 8 and 16 top-widths, we obtained
results consistent within statistical errors of less than 1\%.  
For $t\bar tj$ one can explicitly take the narrow top-width limit and 
compare with the results in Ref.~\cite{RZ_WW}, for example.  
The programs passed all tests.

In order to achieve an accuracy of 1\% in a practical time, the programs use
an enhanced version\footnote{
  The code is available at http://hepsource.org/dvegas/.}
of the VEGAS algorithm \cite{VEGAS1,VEGAS2}.  
It applies importance sampling also to the summation of physical 
helicity combinations and separately optimizes suitably chosen combinations 
of subprocesses/phase space regions. 

\subsection{Numerical Results}

First numerical results were obtained using CTEQ4L parton distribution 
functions as a default, 
with\footnote{This $\alpha_s(m_Z)$ is based on
  $\Lambda^{(5)}_{QCD}$ as determined in the PDF set fit.}
$\alpha_s(m_Z) = 0.132$\cite{CTEQ4_PDF}. The renormalization and
factorization scales $\mu_{\scriptstyle r,f}$ are fixed to the top
mass, $m_t = 175$ GeV.  
In contrast to the
Tevatron, top production at the LHC is dominated by the gluon-gluon channel
and hence noticeably affected by uncertainties of the gluon density.  To
assess the impact of effects related to PDF and $\alpha_s(m_Z)$ choice, we
compared CTEQ4L and CTEQ5L parton distribution functions \cite{CTEQ5_PDF}.
Studying the basic process $pp\to t\bar t$, one obtains 622 and 510 pb for
CTEQ4L and $\alpha_s(m_Z) = 0.132$ and 0.118, respectively.  On the other 
hand, for CTEQ5L and $\alpha_s(m_Z) = 0.127$ and 0.118 one gets 
560 and 487 pb, respectively, suggesting an overall uncertainty of about 
10 or 20\% related to the choice of PDF and $\alpha_s(m_Z)$ related to it.  
With selection cuts, we found similar deviations.  
For example, in Table~\ref{tab:tau_data} below, switching from CTEQ4L
to CTEQ5L leads to a 4-8\% decrease in cross sections at the forward
tagging cuts level.

\begin{table}
\caption{
Comparison of inclusive cross sections with and without finite top
width effects.  All results are given in pb.  The first four columns 
represent our results: full $t\bar t$ cross 
section including off shell effects, contribution within 
$|m_{bW^+}-m_t| > n_c\Gamma_t$ or $|m_{\bar bW^-}-m_t| > n_c\Gamma_t$
with $n_c=$ 10 and 15, 
difference to the NWA result, and this 
difference as a fraction of the total. The $Wt$ column uses 
results from Table~1 in Ref.~\protect\cite{Boos_SingleTop_Followup} for 
$pp\to b\bar bWW$
and Table~1 in Ref.~\protect\cite{Trefzger_Higgs} for $pp\to b\bar
b\ell\bar\nu\bar\ell\nu$.  The results in the second row include
leptonic $\tau$ decays, but not $b$ decays.  
In Ref.~\protect\cite{Boos_SingleTop_Followup} different methods are
used for $Wt$ to reduce overlap with $t\bar t$: $\ \ ^{\ast)}\ |m_{bW}-m_t|
> 10\Gamma_t \ {\rm cut}\quad ^{\S)}\ |m_{bW}-m_t| > 15\Gamma_t$ cut.
}
\vspace{0.15in}
\label{tab:incl_data}
\begin{tabular}{l|cccc|cc}
process & full ${\cal O}(\alpha_s^2\alpha^{2,4})$ & $|m_{bW}-m_t|>10/15\Gamma_t$ & 
full - NWA & rel.contr. &   $W^-t+W^+\bar t$ & rel.contr. \\
\hline
$pp\to b\bar bWW$ & 622 & 63.9/49.6 & 25 & $+4\%$ & $66.4^\ast/51.6^\S$ &
$+11.1/8.6\%$  \\
$pp\to b\bar b\ell\bar\nu\bar\ell\nu$ & 39.2 & 4.0/3.1 & 1.4 & +4\% 
& 4.0 & +10\%
\end{tabular}
\end{table}

A first comparison of our full calculation with single top production 
cross sections 
in the literature is presented in Table~\ref{tab:incl_data}. The two rows 
correspond to $b\bar bWW$ and $b\bar b\ell^+\nu\ell^-\bar\nu$ final states,
i.e. in the second row off-shell-$W$ corrections are included in our 
calculation. In our simulation, $b\bar be^+\nu\mu^-\bar\nu$ results were 
multiplied by a factor 5.48 to account for all lepton flavor combinations,
including leptonic $\tau^\pm$ decays. Off-shell contributions from 
$W^-t+W^+\bar t$ final states were calculated in 
Ref.~\cite{Boos_SingleTop_Followup} with several definitions of the off-shell
region. These results agree with ours at the 4\% level 
(see second column). The agreement improves to the 1--2\% level
once double counting effects in the non-resonant regions ($N$-regions of
Fig.~\ref{fig:phasespace}) are taken into account. Compared to this 
$W^-t+W^+\bar t$ cross section, however, the actual enhancement of 
the $t\bar t$ 
cross section due to off-shell contributions, calculated as the difference 
of our full result minus the cross section in the NWA (see column three), is 
only about half as large.

The problem can be traced to the fact that a Breit-Wigner distribution 
has long tails, more precisely
\bq
\int_{(m - n \Gamma)^2}^{(m + n \Gamma)^2} \frac{1}{(q^2-m^2)^2 + (m\Gamma)^2}
\cdot \frac{dq^2}{2\pi} \approx \frac{1}{2m\Gamma} \cdot \left(1 -
\frac{1}{n\pi}\right)\,
\eq
i.e. for each of the two top-quark resonances, about 2\% of the cross section 
is located outside an $n=15$ top-widths window, for example. This 4\% resonant 
contribution, which is double-counted when combining the rate in the
$|m_{bW}-m_t| > 15\Gamma_t$ region with the $t\bar t$
cross section in the narrow width approximation, accounts for the difference 
between our ``full minus NWA'' result and the estimate in terms
of the $W^-t+W^+\bar t$ cross section. This means that off-shell
$t\bar t$ calculations which directly take $W^-t+W^+\bar t$ cross sections
as the off-shell rate, tend to produce a serious overestimate of the number
of extra events.

Our method of choice to include finite-width effects involves overall
factors that are expected to be small in non-resonant regions as
explained in Section~\ref{sec:method}.  In order to test this expectation
in a realistic context, we compared the differential cross section
obtained with overall factors with various proxies in different phase
space regions (see Fig.~\ref{fig:phasespace}).  In the non-resonant
region $N$ the proxy is the tree-level matrix element with unmodified
tree-level top quark propagators, because they are not singular and a
good approximation to the full propagator in this region.  In the
single-resonant regions $S_t$ and $S_{\bar t}$ the proxy is derived
from the proxy used in region $N$ by replacing the potentially
singular top (in $S_t$) or anti-top (in $S_{\bar t}$) propagators with
fixed-width propagators, i.e.
\bq \frac{i(\sla{p}+m_t)}{p^2-m_t^2+im_t\Gamma_t} \quad .
\label{eq:fixed-width-prop}
\eq

\begin{figure}[htb]
\vspace{0.5cm}
\begin{center}
\includegraphics[width=9.0cm, angle=90]{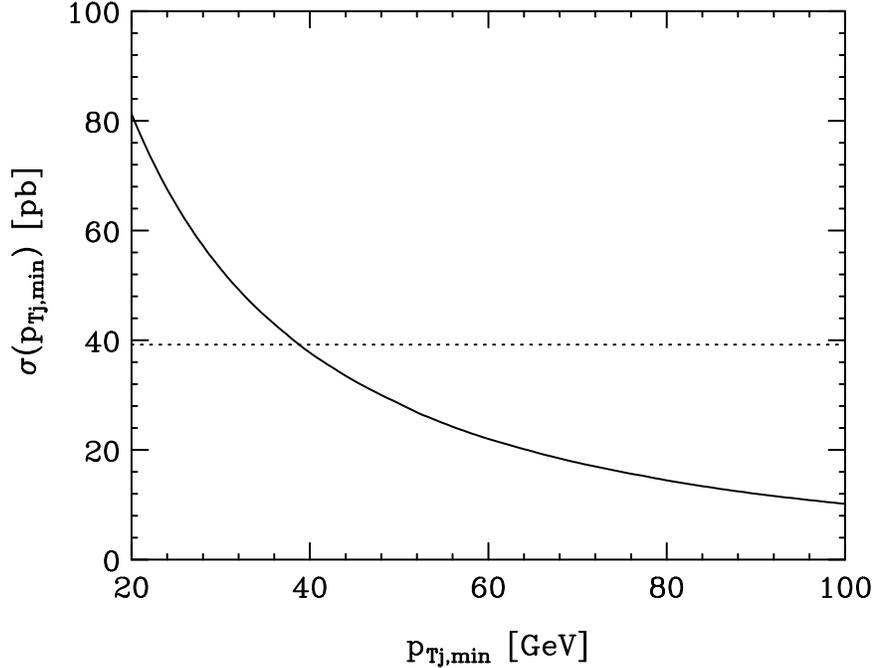} \\
\end{center}
\caption{Integrated transverse momentum distribution for the additional 
massless parton (jet) in off-shell $t\bar tj$ production: 
$\sigma(p_{Tj}^{\text{min}}) =\int_{p_{Tj}^{\text{min}}}^{\infty}
(d\sigma/dp_{Tj})\,dp_{Tj}$.  No selection cuts are
applied.  For $p_{Tj}^{\rm min} = 40$ GeV one obtains $\sigma_{t\bar
tj} \approx \sigma_{t\bar t} = 39.2$ pb (dotted line, see 
Table~\protect\ref{tab:incl_data}).}
\vspace{0.5cm}
\label{fig:ptj_dist}
\end{figure}

In the double-resonant region $D$ the proxy matrix element is a subset
of all Feynman diagrams, namely all diagrams with at least one time-like top
propagator.  Since all these top propagators are potentially resonant here,
they all feature the fixed-width form of Eq.~(\ref{eq:fixed-width-prop}).
Each region is covered by 20-28 bins.  Bin sizes are adjusted so that
each value roughly has the same integration error.
For each bin the relative deviation from the proxy is estimated by
\bq \left|\int_{\scriptstyle \rm bin}
d\sigma_{\scriptstyle \rm fac} -
\int_{\scriptstyle \rm bin} d\sigma_{\scriptstyle
\rm proxy}\right| \;\left/\; \int_{\scriptstyle \rm bin} 
d\sigma_{\scriptstyle \rm proxy} \right. \quad .
\label{eq:proxy-deviation}
\eq
We studied this measure for $t\bar t$ production without cuts as well
as the forward tagging cuts of Refs.~\cite{RZ_WW} and \cite{PRZ_TauTau}
and found similar results.  In the double- and single-resonant regions
the overall factor scheme deviates from the proxies by 2\% or less, and
the deviation drops to less than 1\% in the non-resonant region.
We therefore estimate the error associated with our finite-width
scheme to be of ${\cal O}$(1\%), which is comparable to the
statistical error of our results, but completely negligible compared to 
missing higher order QCD corrections. These results suggest that our method to
include finite-width effects provides reliable results, not only for fairly
inclusive cross sections, where finite width effects are strongly suppressed,
but also when complex selection cuts result in on- and off-shell contributions 
of similar size. 

Comparing inclusive cross sections, we find $B\sigma_{t\bar t} = 39.2$~pb 
in Table~\ref{tab:incl_data}. Including an additional jet we obtain
$B\sigma_{t\bar tj}=80$~pb, when a common $p_T > 20$~GeV cut is 
imposed on the additional jet. This raises the question, whether
the cut should typically be chosen higher, such that 
$\sigma_{t\bar tj} \lesssim \sigma_{t\bar t}$, i.e.~$\sigma_{t\bar tj}$ 
is sub-dominant,
as typically expected for a NLO QCD correction.  The integrated $p_{Tj}$ 
distribution of the non-$b$ parton 
is shown in Fig.~\ref{fig:ptj_dist}.  For $p_{Tj}^{\rm min} = 40$ GeV
one obtains $\sigma_{t\bar tj} \approx \sigma_{t\bar t}$. Real parton emission 
cross sections saturating the LO cross section at low $p_T$ of the extra 
parton are a sign of copious multi-jet production in actual data~\cite{rsz}.

\begin{figure}[htb]
\vspace*{0.5cm}
\begin{center}
\begin{minipage}[c]{.48\linewidth}
\flushright \includegraphics[width=7.0cm, angle=90]{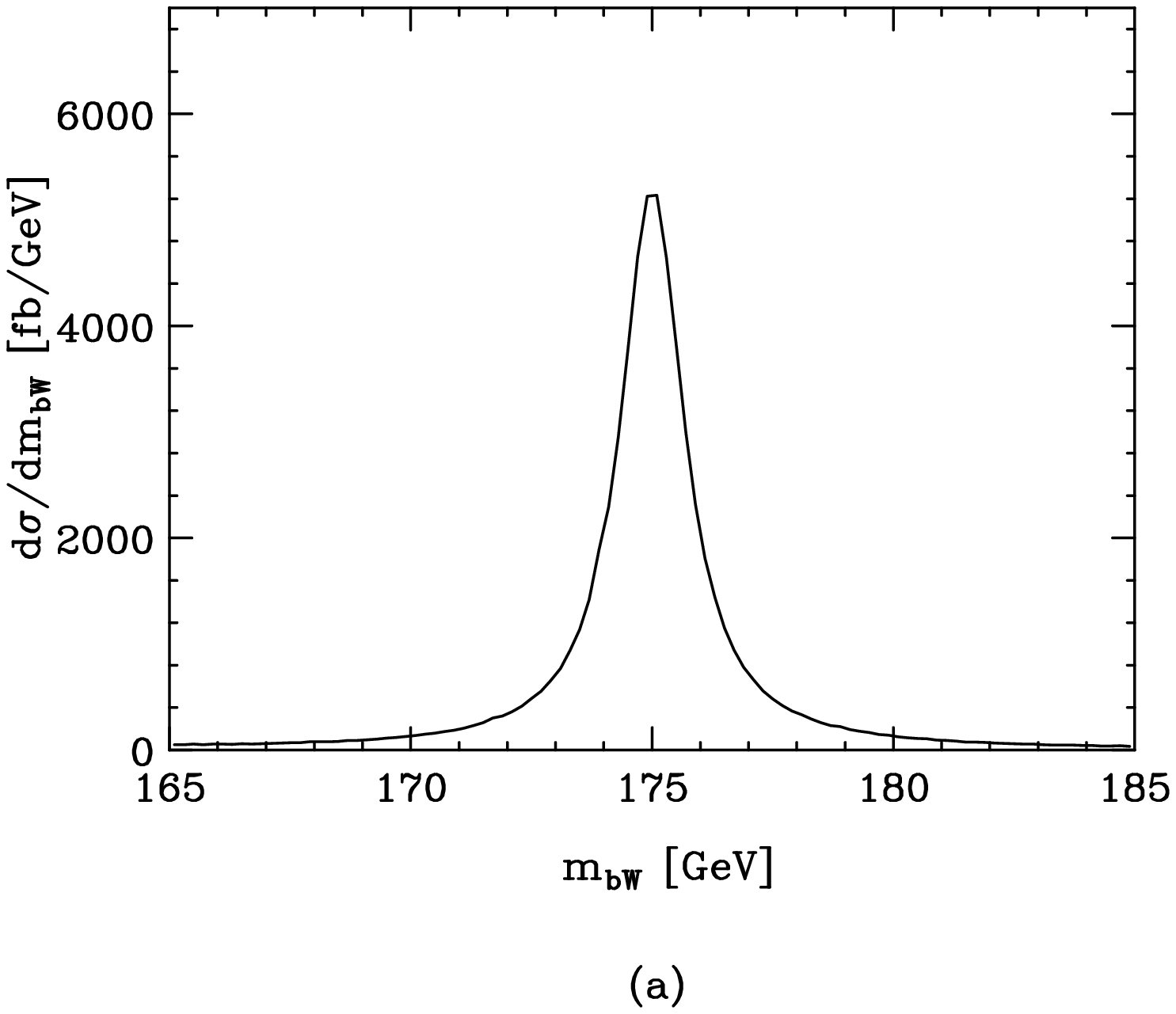} 
\end{minipage} \hfill
\begin{minipage}[c]{.48\linewidth}
\flushleft \includegraphics[width=7.0cm, angle=90]{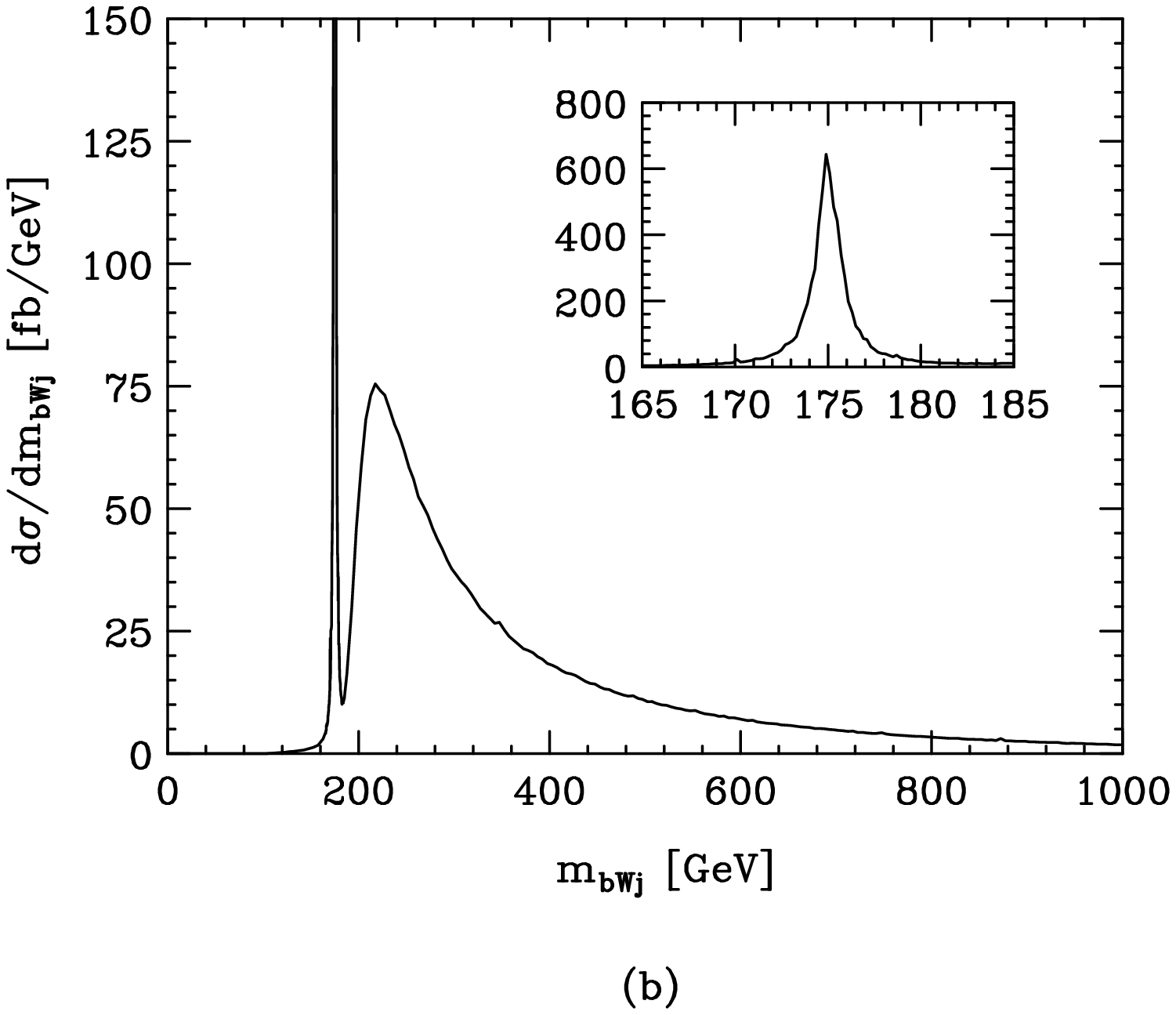} 
\end{minipage}
\end{center}
\caption{Invariant mass distribution of potential top-quark decay products
(a) $bW^+$ and (b) $bW^+j$ in $t\bar tj$ production, including full
finite width effects.  No selection cuts are applied, except 
for a $p_{Tj} > 20$~GeV cut on the final state massless parton.  
Cross sections correspond to one lepton-flavor combination only.  
}
\vspace{0.5cm}
\label{fig:bW_mass}
\end{figure}

The additional parton emission is dominated by initial state radiation.
This can be inferred from the invariant mass distributions of the potential
top-quark decay products, the $bW^+$ and the $bW^+j$ systems shown in 
Fig.~\ref{fig:bW_mass}. The $m_{bW^+}$ invariant mass distribution in 
Fig.~\ref{fig:bW_mass}(a) contains 85\% of the total cross section in 
the displayed 165-185 GeV window around the top resonance ($\pm6.4\Gamma_t$). 
In contrast, the same $m_{bW^+j}$ invariant mass window 
(insert of Fig.~\ref{fig:bW_mass}(b)) accounts for only 10\% of the total 
cross section. Final state radiation is relatively unimportant in $t\bar tj$
production at the LHC.

\section{Application to SM Higgs search at the LHC}
\label{sec:application}

\begin{table}[t]
\caption{
Numbers of expected $pp\to b\bar b\ell\bar\nu\bar\ell\nu$ background events
in the ATLAS $H\to WW^*$ Higgs search for 
an integrated luminosity of 30 fb$^{-1}$. Basic ATLAS TDR
background suppression cuts are applied
\protect\cite{ATLAS_TDR_2,Trefzger_Higgs}. ATLAS detector 
effects are simulated as described in Section~II.F in
Ref.~\protect\cite{RZ_WW}.  The left columns show our results for the  
four $W\to e\nu,\mu\nu$ lepton combinations. The right
columns are the corresponding results from Table~2 in
Ref.~\protect\cite{Trefzger_Higgs}. For both categories, the relative 
change due to the inclusion of finite width effects is given as an 
enhancement factor in the third column.}
\vspace{0.15in}
\label{tab:tdr_data}
\begin{tabular}{l|ccc|ccc}
$m_T$ window [GeV] & top in NWA & full off-shell & rel.~chg. & 
$t\bar{t}$ & $t\bar{t}+Wt$ & rel.~chg. \\
\hline
120--150 & 50 &       93 &     1.9 & 41 & 206 &  5.0  \\
130--160 & 50 &       96 &     1.9 & 66 & 204 &  3.1  \\
140--170 & 43 &       83 &     1.9 & 51 & 146 &  2.9  \\
140--180 & 49 &       95 &     1.9 & 61 & 164 &  2.7  \\
150--190 & 35 &       68 &     1.9 & 56 & 111 &  2.0  
\end{tabular}
\end{table}

An important application of $t\bar t{j}$ production as a background 
occurs in Higgs physics.
Higgs mass limits have recently been pushed to $m_H>114.1$ GeV by the LEP 
experiments~\cite{lep2higgs}. As a result, the LHC search for 
$H\to WW$~\cite{DittDrein,CMS,ATLAS_TDR_2,Trefzger_Higgs,KPRZ,RZ_WW}  and 
$H\to\tau\tau$~\cite{ATLAS_TDR_2,RZH_tau,PRZ_TauTau} decays in the 
intermediate mass range has gained even greater importance. For the 
$H\to WW\to \ell^\pm\ell^\mp\sla{p}_T$ decay mode, top-quark decays 
constitute the largest reducible background.  The impact of this 
background has been analyzed, in the narrow-width approximation, for 
Higgs masses around 170~GeV~\cite{ATLAS_TDR_2,Trefzger_Higgs,RZ_WW} and 
most recently for a light Higgs boson with 
$m_H\approx 115$~GeV~\cite{KPRZ}.
This background also plays a role in the $H\to\tau\tau\to
e^\pm\mu^\mp\sla{p}_T$ decay mode, which was analyzed in the narrow-width
approximation in Ref.~\cite{PRZ_TauTau}.  In this section we present
updated results for these analyses obtained with our parton-level Monte
Carlo program that includes off-shell top and $W$ effects and takes into
account the single-resonant and non-resonant contributions.\footnote{The
  analysis in Ref.~\protect\cite{KPRZ} already features
  off-shell $t\bar t$ and $t\bar tj$ background estimates calculated with
  the programs presented here.}

\subsection{\lowercase{$t\bar t$} Backgrounds to Inclusive $H\to WW$ Searches}

The $t\bar t$ background calculations (without an additional jet) are most 
relevant for inclusive
$H\to WW$ searches~\cite{DittDrein,CMS,ATLAS_TDR_2}, where Higgs production 
is dominated by the gluon fusion process. In Tables~\ref{tab:tdr_data} and
\ref{tab:opt_data} we compare the relative contributions from off resonant 
top effects for two selections of $H\to WW\to \ell^\pm\ell^\mp\sla{p}_T$
events in ATLAS, as described in Refs.~\cite{ATLAS_TDR_2,Trefzger_Higgs}.
The selection looks for two isolated, opposite charge leptons of
$p_T>20,10$~GeV within the pseudo-rapidity range $|\eta_\ell|<2.5$
and of invariant mass $m_{\ell\ell}<80$~GeV. Events must have significant 
missing $E_T$, $\sla E_T>40$~GeV. A small angle between the charged leptons
favors $H\to WW$ decays versus backgrounds. Finally, a veto on additional 
jets in the central region is very effective against the $b$-quark jets 
in the top-quark backgrounds. The main difference between 
Tables~\ref{tab:tdr_data} and \ref{tab:opt_data} is the definition of this 
veto on jet activity with $p_T>15$~GeV in the central region. It is imposed 
within $|\eta_j|<3.2$ in Table~\ref{tab:tdr_data} and within $|\eta_j|<2.4$ 
in Table~\ref{tab:opt_data}.

\begin{figure}[htb]
\vspace{0.5cm}
\begin{center}
\includegraphics[width=9.0cm, angle=90]{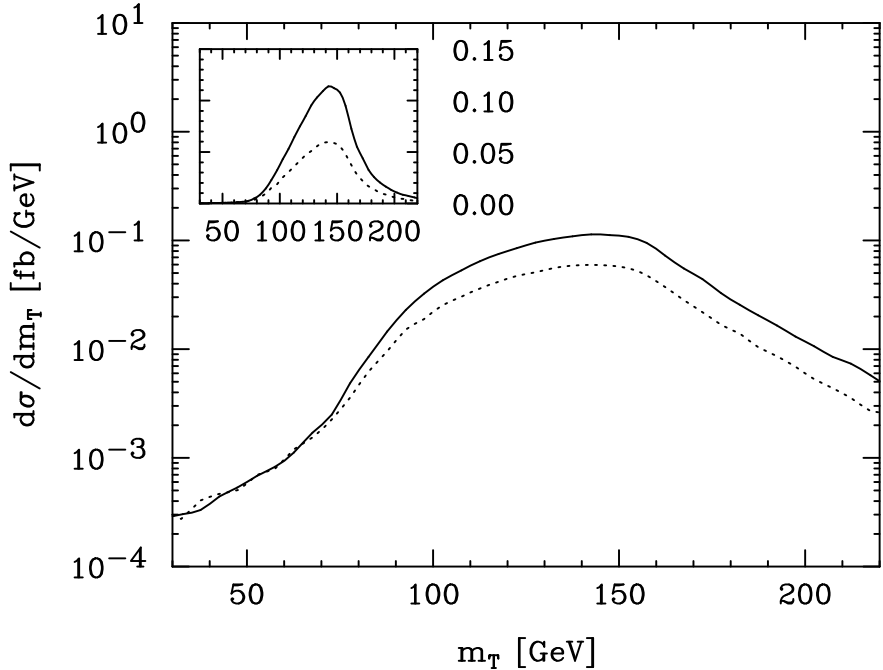} \\
\end{center}
\caption{Transverse mass distribution for the low-luminosity ATLAS selection
cuts of Ref.~\protect\cite{ATLAS_TDR_2}, Section~19.2.6, and
Ref.~\protect\cite{Trefzger_Higgs}.  The dotted curve shows the distribution
obtained when treating the top quarks in the narrow-width approximation
while the solid curve includes all off-shell effects. }
\vspace{0.5cm}
\label{fig:mT}
\end{figure}

In addition, the selection looks for events inside the Jacobian peak 
of the dilepton-$\sla E_T$ transverse mass distribution, as indicated in the 
first column of the two tables. Here the transverse mass is defined as
\bq
m_T(\ell\ell,\sla E_T) = 
\sqrt{2p_T^{\ell\ell}\sla E_T[1-\cos\Delta\theta(\ell\ell,\sla E_T)]}\;.
\label{eq:mtrans}
\eq
Fig.~\ref{fig:mT} shows this transverse mass distribution underlying
the ATLAS TDR analysis.  Off-shell contributions raise the normalization
of the $t\bar t$ background by about a factor of 2, but have little effect
on the shape of the background, for $m_T \gtrsim 120$ GeV. The ATLAS
analysis attempts to take (single-resonant) off-shell effects into
account via on-shell $Wt$ calculations.  As the comparison in 
Table~\ref{tab:tdr_data} shows, our unified calculation indicates a
lower background increase than the combined on-shell $t\bar t$ and $Wt$
calculations would suggest. This observation is consistent with the 
small increase due to off-resonant effects observed in 
Table~\ref{tab:incl_data}.  

For a precise comparison with the ATLAS
simulation, the nature of the programs (parton-level MC with full matrix 
elements vs.~event generator with parton showers/hadronization) is currently 
too different. In particular our program does not allow to simulate 
the effect of the central jet veto on extra gluon radiation in the event.
Also, we did not include any detector efficiencies. 
We would expect an extra suppression, by perhaps a factor of 3 to 4, of the 
``full off-shell'' results in Tables~\ref{tab:tdr_data} and
\ref{tab:opt_data} due to these effects, i.e. the apparent agreement of 
the $t\bar t$ cross 
section with our NWA result in Table~\ref{tab:tdr_data} is somewhat 
fortuitous. This suspicion is confirmed by the fairly large disagreement
between our NWA or full off-shell results  and the PYTHIA simulation in 
Table~\ref{tab:opt_data}. 

\begin{table}[t]
\caption{
Same as Table~\protect\ref{tab:tdr_data}, but for the optimized background
suppression cuts of Ref.~\protect\cite{Trefzger_Higgs}. }
\vspace{0.15in}
\label{tab:opt_data}
\begin{tabular}{l|ccc|ccc}
$m_T$ window [GeV] & top in NWA & full off-shell & rel.~chg. & 
$t\bar{t}$ & $t\bar{t}+Wt$ & rel.~chg. \\
\hline
120--150 & 364 &       536 &     1.5 & 107 & 346 &  3.2  \\
130--160 & 363 &       544 &     1.5 & 122 & 306 &  2.5  \\
140--170 & 310 &       465 &     1.5 &  82 & 215 &  2.6  \\
140--180 & 352 &       528 &     1.5 & 107 & 254 &  2.4  \\
150--190 & 251 &       374 &     1.5 &  87 & 168 &  1.9  
\end{tabular}
\end{table}

At present the source of these discrepancies is not completely understood.
The rapidity distribution of $b$-quarks in our full matrix element 
calculation appears to be wider than in the PYTHIA generated events,
potentially due to approximations in the top decay chain in PYTHIA.
This makes a veto over a smaller pseudo-rapidity range less efficient. 
On the other hand we cannot simulate the effect of additional 
gluon radiation. A reanalysis of these effects, combining full matrix
elements with a parton shower Monte Carlo, including hadronization, is clearly 
warranted, given that top-quark production constitutes about half the
background to the inclusive $H\to WW$ search\cite{Trefzger_Higgs}.

While the overall background normalization requires further study, the
ratio of ``full off-shell'' and NWA results is expected to be robust, i.e.
it will be little affected by detector efficiencies and by higher order
gluon radiation. These ratios are given in the third columns, marked 
``relative change'', in Tables~\ref{tab:tdr_data} and \ref{tab:opt_data}.
Our results clearly indicate that the increase in 
top-backgrounds due to the inclusion of off-shell effects (the $Wt$ 
contribution in the ATLAS analysis) is substantially smaller than previously
thought. 

In a LO calculation, substantial uncertainties arise from ambiguities
in the choice of factorization and renormalization scales.
For the double resonant phase space configurations, 
$\mu_{\scriptstyle r}=\mu_{\scriptstyle f}=m_t$ or
the transverse energy of the top quarks are well-motivated choices.  
When forcing the $b$-quarks to low $p_T$ values by a central jet veto
and simultaneously enhancing the off-shell phase space regions, a smaller
renormalization and/or factorization scale may be more appropriate.
The impact of a lower scale on the full off-shell results 
in Table~\ref{tab:tdr_data} is demonstrated by the scale choice
\bq \mu_{\scriptstyle r,f} = 
\left\{ \begin{array} {r@{\quad\mbox{if}\quad}l} 
m_t & |m_{bW^+}-m_t| < 4\Gamma_t \;\, \wedge \;\, 
|m_{\bar bW^-}-m_t| < 4\Gamma_t \\
20\ \mbox{GeV} & |m_{bW^+}-m_t| > 4\Gamma_t \;\, \vee \;\, 
|m_{\bar bW^-}-m_t| > 4\Gamma_t
\end{array}\right. 
\label{eq:lower-scale}
\eq
where the value of 20~GeV is motivated by the veto threshold for 
central jets.
Resulting cross sections are about 65\% higher than the rates obtained
with $m_t$ as a universal scale.  A NLO calculation would be needed
to distinguish the virtue of either choice. At present, this variation
indicates the uncertainty of our LO results.

\subsection{\lowercase{$t\bar t(j)$} Backgrounds in Weak Boson Fusion Studies}

In addition to the inclusive search for $H\to WW$ events, weak boson fusion
(WBF) presents a very attractive search channel for $H\to WW$ and 
$H\to\tau\tau$
events and will likely play a crucial role in the measurement of Higgs
boson couplings to fermions and gauge bosons~\cite{knrz}. Top quark decays
again form an important background in these searches and, due to the central
jet veto proposed for the reduction of QCD backgrounds, off-shell effects 
might be important. The signal and backgrounds for 
$H\to WW\to e^\pm\mu^\mp\sla{p}_T$ and $H\to\tau\tau\to e^\pm\mu^\mp\sla{p}_T$
in WBF were analyzed in Refs.~\cite{RZ_WW} and \cite{PRZ_TauTau}, 
respectively, with the top-quark backgrounds determined in the NWA, however. 
With our new programs we are able to update these background estimates,
including off-shell contributions. 

Because of the two additional jets which are present in the WBF process
$qq\to qqH$, the dominant top quark background arises from $t\bar tj$ events.
Event selection for WBF requires two tagging jets, of $p_T>20$~GeV, which are
widely separated in pseudo-rapidity ($|\eta_1 - \eta_2|>4.2$) and which have 
a very large dijet invariant mass, $M_{jj}>650$~GeV. By definition,
the $t\bar t$ background has both $b$-quarks identified as tagging jets 
while in the $t\bar tj$ background exactly one $b$ or $\bar b$ is
taken as a tagging jet. The two $b$-quarks rarely have a large enough
dijet mass or are far enough separated to satisfy the tagging criteria,
and this leaves $t\bar tj$ events as the dominant background to WBF. 
This result holds in both the NWA and with inclusion of off-resonant effects
as is evident in both Tables~\ref{tab:WW_data} and \ref{tab:tau_data},
at all cut levels.

The veto of central jets of $p_T>20$~GeV is effective against the extra 
$b$-quark jet and off-shell contributions are only slightly enhanced 
after this cut (see line ``$b$ veto'' in the Tables).  Overall,
off-shell contributions are fairly modest, increasing the NWA results 
by about 20\% (see the second last columns in the Tables). This means 
that our new complete calculation of off-shell effects in $t\bar tj$ 
production confirms the conclusions about the observability of $H\to\tau\tau$
and $H\to WW$ events reached in Refs.~\cite{RZ_WW} and \cite{PRZ_TauTau}.
Tables~\ref{tab:WW_data} and \ref{tab:tau_data} display our updated results.
Precise definitions of cuts are given in the earlier papers. A breakdown
into subprocesses and phase space regions of the overall $t\bar tj$ 
background of 351 fb to $H\to WW\to e^\pm\mu^\mp\sla{p}_T$, after forward 
jet tagging cuts, is given in Table~\ref{tab:WWbreakdown}.

\begin{table}
\caption{
$t\bar{t}(j)$ background cross sections for 
$H\to WW\to e^\pm\mu^\mp\sla{p}_T$ for 
$m_H = 160$~GeV in $pp$ collisions at $\protect\sqrt{s}=14$~TeV. 
Results are given for 
various levels of cuts and are labeled by equation numbers from 
Ref.~\protect\cite{RZ_WW}.  
All cross sections are given in fb.  Cuts and other calculational
details are described in Ref.~\protect\cite{RZ_WW}.  The integration
error is 1\% or better.  The signal over background ratio is also
shown.  Cross sections not listed here are as in Table~I in 
Ref.~\protect\cite{RZ_WW}.}
\vspace{0.15in}
\label{tab:WW_data}
\begin{tabular}{l|ccc|ccccc}
& \multicolumn{3}{c|}{top in NWA} & \multicolumn{5}{c}{full off-shell} \\
cuts & $t\bar{t}$ & $t\bar{t}j$ & S/B & 
\multicolumn{2}{c}{$t\bar{t}$} & \multicolumn{2}{c}{$t\bar{t}j$} & S/B \\
\hline
forward tagging (10)-(12)
& 12.4 & 308 & $\approx$1/65 & 13.0 & +4.4\% & 351 & +14\% & $\approx$1/67 \\
+ $b$ veto (13)
& & 43.5 & 1/5.1 &  & & 51.4 & +18\% & 1/5.6 \\
+ $M_{jj}$, angular cuts (14)-(16)
& 0.0551 & 4.67 & 1.1/1 & 0.0761 & +38\% & 5.42 & +16\% & 1.0/1 \\
+ real $\tau$ rejection (17)
& 0.0527 & 4.34 & 1.7/1 & 0.0737 & +40\% & 5.09 & +17\% & 1.5/1 \\
$P_{surv,20}$ (${\it\times 0.29}$) + (18)
& 0.0153 & 1.26 & 4.6/1 & 0.0214 & +40\% & 1.48 & +17\% & 4.2/1 \\
+ tag ID efficiency (${\it\times 0.74}$)
& 0.0113 & 0.932 & 4.6/1 & 0.0158 & +40\% & 1.09 & +17\% & 4.2/1 \\
\end{tabular}
\end{table}

\begin{table}
\caption{
$t\bar{t}(j)$ background cross sections for 
$H\to\tau\tau\to e^\pm\mu^\mp\sla{p}_T$ for 
$M_H = 120$~GeV in $pp$ collisions at $\protect\sqrt{s}=14$~TeV. 
Results are given for 
various levels of cuts and are labeled by equation numbers from 
Ref.~\protect\cite{PRZ_TauTau}.  
All cross sections are given in fb.  Cuts and other calculational
details are described in Ref.~\protect\cite{PRZ_TauTau}.  The
integration error is 1\% or better. The signal over background ratio is also
shown.  Cross sections not listed here are as in Table~I in
Ref.~\protect\cite{PRZ_TauTau}.}
\vspace{0.15in}
\label{tab:tau_data}
\begin{tabular}{l|ccc|ccccc}
&  \multicolumn{3}{c|}{top in NWA} & \multicolumn{5}{c}{full off-shell} \\
cuts & $t\bar{t}$ & $t\bar{t}j$ & S/B & 
\multicolumn{2}{c}{$t\bar{t}$} & \multicolumn{2}{c}{$t\bar{t}j$} & S/B \\
\hline
forward tagging (7)-(10)
& 13.5 & 357 & 1/1100 & 15.9 & +17\% & 436 & +22\% & 1/1100 \\
+ $b$ veto (11)
& & 50.1 & 1/550 & & & 63.6 & +27\% & 1/550 \\
+ $\sla{p}_T$ (12)
& 11.1 & 43.0 & 1/74 & 13.2 & +19\% & 55.2 & +28\% & 1/83 \\
+ $M_{jj}$ (13)
& 0.593 & 12.9 & 1/32 & 0.712 & +20\% & 15.8 & +22\% & 1/34 \\
+ non-$\tau$ reject. (14, 15, 17)
& 0.00303 & 0.257 & 1/5.8 & 0.00365 & +20\% & 0.293 & +14\% & 1/5.8 \\
\end{tabular}
\end{table}

\begin{table}
\caption{Distribution of the $t\bar tj$ background to $H\to WW\to
e^\pm\mu^\mp\sla{p}_T$, with forward tagging cuts, among subprocesses 
(labeled by initial partons) and phase space regions 
(see Fig.~\protect\ref{fig:phasespace}). The cut between double resonant ($D$),
single resonant ($S_t+S_{\bar t}$), and non-resonant regions ($N$) is
set at 8 top quark widths. Cross sections are given in fb.}
\vspace{0.15in}
\label{tab:WWbreakdown}
\begin{tabular}{l|ccc}
         & $D$  & $S_t+S_{\bar t}$ & $N$ \\
\hline
$gg$     & 171  &  28.6            & 1.1 \\
$gq+qg$  & 128  &  18.8            & 0.64 \\
$qq$     & 2.92 &   0.35           & 0.009 \\
\end{tabular}
\end{table}


\section{Summary}
\label{sec:summary}

Top-quarks are a very important source of lepton backgrounds to new physics 
searches at the LHC: the large production cross section typical for a
strong interaction process combines with a sizable branching fraction
into leptons which, due to the large mass of the $W$, often survive 
isolation cuts. Suppression techniques for top quark backgrounds, like
a veto on central jets, which is very effective against the $b$-quarks
of the $t\to bW\to b\ell\nu$ decay chain, enhance the relative importance 
of off-resonant effects and may exacerbate errors introduced by  
approximate modeling of matrix elements. Severe cuts select the tails of 
various distributions and these phase space regions may well differ from 
the ones for which the models were optimized 
originally. General purpose Monte Carlo programs like PYTHIA\cite{pythia}
or Herwig\cite{herwig}
should thus be gauged against matrix element programs.

In this paper we have presented results for two new programs which allow
to model the  $t\bar t\to b\bar bWW \to b\bar b \ell^+\nu\ell^-\bar\nu$
decay chain at tree level, including full angular correlations of all
top and $W$ decay products and with proper interpolation between 
double-resonant, single-resonant and non-resonant phase space regions. 
These full correlations are available for $t\bar t$ and $t\bar tj$ production.
Electroweak and $SU(3)$ gauge invariance is maintained throughout, by employing
the overall factor scheme for the Breit Wigner propagators of
all unstable particles in the $t$ and $\bar t$ decay chains.

Comparing to earlier calculations of off-shell effects, via the $gb\to Wtb$ 
production cross sections, we find excellent agreement when avoiding the 
top-quark resonance for the $Wb$ system.  However, some earlier combinations 
of $t\bar t$ and $Wtb$ cross sections have involved substantial
double counting, leading to an overestimate of backgrounds in e.g. 
Higgs search analyses at the LHC. In addition, the full simulation of $V-A$
couplings in the $t\to Wb\to \ell\nu b$ decay chain is available with our
programs and may have sizable effects in background estimates.

When considering $t\bar t$ backgrounds to Higgs searches, off-shell effects 
are most pronounced in inclusive $H\to WW$ analyses, where both $b$-quark 
jets may be vetoed. In weak boson fusion studies the additional jets
in the signal selection make the background suppression due to the jet veto
less severe, which diminishes the overall importance of off-shell 
contributions. 

Our programs now allow detailed study of these off-shell effects in top
pair backgrounds with zero or one additional parton in the final state.


\acknowledgements 
We thank K.~Jakobs, T.~Trefzger and T.~Plehn for useful discussions.  This research was
supported in part by the University of Wisconsin Research Committee
with funds granted by the Wisconsin Alumni Research Foundation and in
part by the U.~S.~Department of Energy under Contract
No.~DE-FG02-95ER40896. 



\begin{references}

\bibitem{CDFD0}
F. Abe {\it et~al.} [CDF Collaboration], Phys. Rev. Lett. {\bf 74}, 2626 (1995); 
S. Abachi {\it et~al.} [D0 Collaboration], Phys. Rev. Lett. {\bf 74}, 2632 (1995).

\bibitem{SUSY}
S. Abel {\it et~al.}, SUGRA Working Group Report, (2000) [hep-ph/0003154];
H. Baer, J.~K. Mizukoshi, and X. Tata, Phys. Lett. {\bf B488}, 367 (2000);
H. Baer, P.~G. Mercadante, X. Tata, and Y. Wang, Phys. Rev. {\bf D62}, 095007 (2000);
V. Barger and C. Kao, Phys. Rev. {\bf D60}, 115015 (1999);
K.~T. Matchev and D.~M. Pierce, Phys. Rev. {\bf D60}, 075004 (1999);
H. Baer {\it et~al.}, Phys. Rev. {\bf D61}, 095007 (2000);
W. Beenakker {\it et~al.}, Phys. Rev. Lett. {\bf 83}, 3780 (1999);
H.~E. Haber and G.~L. Kane, Phys. Rept. {\bf 117}, 75 (1985).

\bibitem{DittDrein}
M.~Dittmar and H.~Dreiner, 
Phys.\ Rev.\ {\bf D55}, 167 (1997); 
and [hep-ph/9703401].

\bibitem{CMS}
G.~L.~Bayatian {\it et al.}, CMS Technical Proposal,
report CERN-LHCC-94-38, (1994); 
R. Kinnunen and D. Denegri, 
CMS NOTE 1997/057, (1997); 
R. Kinnunen and A. Nikitenko, 
CMS TN/97-106, (1997); 
R. Kinnunen and D. Denegri, 
CMS NOTE 1999/037, (1999).   

\bibitem{ATLAS_TDR_2}
{ATLAS Collaboration}, report CERN-LHCC-99-15,   (1999).

\bibitem{Trefzger_Higgs}
K. Jakobs and T. Trefzger,  note ATL-PHYS-2000-015, (2000).

\bibitem{KPRZ}
N. Kauer, T. Plehn, D. Rainwater, and D. Zeppenfeld, Phys. Lett. {\bf B503},
  113  (2001).

\bibitem{RZ_WW}
D. Rainwater and D. Zeppenfeld, Phys. Rev. {\bf D60},  113004  (1999).

\bibitem{RZH_tau}
D.~Rainwater, D.~Zeppenfeld, and K.~Hagiwara, 
Phys.~Rev.~{\bf D59}, 014037 (1999).

\bibitem{PRZ_TauTau}
T. Plehn, D. Rainwater, and D. Zeppenfeld, Phys. Rev. {\bf D61},  093005
  (2000).

\bibitem{Boos_SingleTop}
A.~S. Belyaev, E.~E. Boos, and L.~V. Dudko, Phys. Rev. {\bf D59},  075001
  (1999).

\bibitem{Tait_tW}
T.~M.~P. Tait, Phys. Rev. {\bf D61},  034001  (2000).

\bibitem{Boos_SingleTop_Followup}
A. Belyaev and E. Boos, Phys. Rev. {\bf D63},  034012  (2001).

\bibitem{OtherSingleTopRefs}
T. Stelzer, Z. Sullivan, and S. Willenbrock, Phys. Rev. {\bf D58},  094021
  (1998);
T. Tait and C.~P. Yuan, Phys. Rev. {\bf D63},  014018  (2001);
M.~C. Smith and S. Willenbrock, Phys. Rev. {\bf D54},  6696  (1996);
S. Mrenna and C.~P. Yuan, Phys. Lett. {\bf B416},  200  (1998);
T. Stelzer, Z. Sullivan, and S. Willenbrock, Phys. Rev. {\bf D56},  5919
  (1997);
T. Tait and C.~P. Yuan, Phys. Rev. {\bf D55}, 7300  (1997).

\bibitem{MADGRAPH}
T. Stelzer and W.~F. Long, Comput. Phys. Commun. {\bf 81},  357  (1994).

\bibitem{HELAS}
H. Murayama, I. Watanabe, and K. Hagiwara, report KEK-91-11,   (1992).

\bibitem{Baur:1995aa}
U. Baur and D. Zeppenfeld, Phys. Rev. Lett. {\bf 75},  1002  (1995).

\bibitem{Argyres:1995ym}
E.~N. Argyres {\it et~al.}, Phys. Lett. {\bf B358},  339  (1995).

\bibitem{Beenakker:1997kn}
W. Beenakker {\it et~al.}, Nucl. Phys. {\bf B500},  255  (1997).

\bibitem{Kauer_Master}
N. Kauer, Master's thesis, University of Wisconsin-Madison, 1996.

\bibitem{Baur:1997bn}
U. Baur {\it et~al.}, Phys. Rev. {\bf D56},  140  (1997).

\bibitem{HPZH}
K.~Hagiwara, K.~Hikasa, R.~D.~Peccei, and D.Zeppenfeld,
Nucl. Phys. {\bf B282}, 253 (1987);
U.~Baur and D.~Zeppenfeld, Phys. Lett. {\bf 201B}, 383 (1988).

\bibitem{bvz}
U.~Baur, J.~Vermaseren, and D.~Zeppenfeld, Nucl.~Phys.~{\bf B375}, 3 (1992). 

\bibitem{Denner:1999gp}
A. Denner, S. Dittmaier, M. Roth, and D. Wackeroth, Nucl. Phys. {\bf B560},  33
   (1999).

\bibitem{compMother}
G.~Lopez Castro, J.~L.~Lucio, and J.~Pestieau,
Mod.~Phys.~Lett.~{\bf A6}, 3679 (1991);
M.~Nowakowski and A.~Pilaftsis,
Z.~Phys.~{\bf C60}, 121 (1993);
M.~Beuthe, R.~Gonzalez Felipe, G.~Lopez Castro, and J.~Pestieau,
Nucl.~Phys.~{\bf B498}, 55 (1997);
G.~Lopez Castro and G.~Toledo Sanchez,
Phys.~Rev.~{\bf D61}, 033007 (2000).

\bibitem{vanderHeide:2000fx}
J. van~der Heide, E. Laenen, L. Phaf, and S. Weinzierl, Phys. Rev. {\bf D62},
  074025  (2000).

\bibitem{VEGAS1}
G.~P. Lepage, J. Comput. Phys. {\bf 27},  192  (1978).

\bibitem{VEGAS2}
G.~P. Lepage, preprint CLNS-80/447,   (1980).

\bibitem{CTEQ4_PDF}
H.~L. Lai {\it et~al.}, Phys. Rev. {\bf D55},  1280  (1997).

\bibitem{CTEQ5_PDF}
H.~L. Lai {\it et~al.}, Eur. Phys. J. {\bf C12},  375  (2000).

\bibitem{rsz}
D.~Rainwater, R.~Szalapski, and D.~Zeppenfeld,
Phys.~Rev.~{\bf D54}, 6680 (1996).

\bibitem{lep2higgs}
ALEPH, DELPHI, L3, and OPAL Collaborations, LHWG NOTE/2001-03, (2001) [hep-ex/0107029].

\bibitem{knrz}
D.~Zeppenfeld, R.~Kinnunen, A.~Nikitenko, and E.~Richter-W\c{a}s,
Phys.~Rev.~{\bf D62}, 013009  (2000).

\bibitem{pythia}
T.~Sjostrand {\it et~al.}, Comput. Phys. Commun. {\bf 135}, 238 (2001).

\bibitem{herwig}
G.~Corcella {\it et~al.}, JHEP {\bf 01}, 010 (2001); and [hep-ph/0107071].

\end{references}
\end{document}